\documentclass[lettersize,journal]{IEEEtran}
\usepackage{amsmath,amsfonts}
\usepackage{algorithmic}
\usepackage{algorithm}
\usepackage{array}
\usepackage[caption=false,font=normalsize,labelfont=sf,textfont=sf]{subfig}
\usepackage{textcomp}
\usepackage{stfloats}
\usepackage{url}
\usepackage{verbatim}
\usepackage{graphicx}
\usepackage{cite}
\usepackage{multirow}

\usepackage{hyperref}
\usepackage{cite}
\usepackage[normalem]{ulem}
\usepackage{soul}
\usepackage{xcolor}
\usepackage{amsmath,amssymb,amsfonts}
\usepackage{pxfonts}
\usepackage{listings}%
\usepackage{multirow}
\usepackage{tikz}

\usepackage{array}
\usepackage{fancyhdr}
\usepackage{arydshln} 
\usepackage{eso-pic}

\newcolumntype{?}{!{\vrule width 0.3mm}}

\makeatletter

\IEEEtriggercmd{\reset@font\normalfont\fontsize{7pt}{8.4pt}\selectfont}
\makeatother
\IEEEtriggeratref{1}
    
\begin{document}
\fancypagestyle{IEEEtitlepagestyle}{%
  \fancyhf{}
  \fancyhead[C]{}
  \renewcommand{\headrulewidth}{0pt}
}
\title{JugglePAC: a Pipelined Accumulation Circuit\\}

\newcommand{\mytikzmark}[1]{%
  \tikz[overlay,remember picture,baseline] \coordinate (#1) at (0,0) {};}

\newcommand{\highlight}[2]{%
  \draw[red,line width=10pt,opacity=0.25]%
    ([yshift=3pt]#1) -- ([yshift=3pt]#2);%
}
\newcommand{\algorithmautorefname}{Algorithm}
\newcommand\Furkancomment[1]{\textcolor{red}{#1}}

\newcommand{\orcid}[1]{\href{https://orcid.org/#1}{\includegraphics[width=10pt]{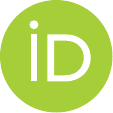}}}

\author{Ahmad Houraniah \orcid{0000-0001-7925-8888},  H. Fatih Ugurdag \orcid{0000-0002-6256-0850} ~\IEEEmembership{(Senior Member, IEEE)} , Furkan Aydin \orcid{0000-0001-7162-4935}
\vspace{-2em}
\thanks{A. Houraniah  is with the Dept. of Computer Science, Ozyegin University, Istanbul 34794, Turkey (email: ahmad.houraniah@ozu.edu.tr).}
\thanks{ H. F. Ugurdag is with the Dept. of Electrical and Electronics Engineering, Ozyegin University, Istanbul 34794, Turkey.
}
\thanks{F. Aydin is with the Dept. of Electrical and Computer Engineering, North Carolina State University, Raleigh, NC 27606, USA.}

}

\markboth{}%
{Houraniah \MakeLowercase{\textit{et al.}}: JugglePAC}
\fancyhf{}
\fancyhead[C]{}
\renewcommand{\headrulewidth}{0pt} 


\maketitle

\newcommand{\ieeetopheader}{%
  \parbox{\textwidth}{\centering\footnotesize
    This article has been accepted for publication in IEEE Embedded Systems Letters. 
    This is the author's accepted version; content may change. DOI: \href{https://doi.org/10.1109/LES.2025.3602852}{10.1109/LES.2025.3602852}
  }%
}

\newcommand{\ieeebottomfooter}{%
  \parbox{1.05\textwidth}{\centering\footnotesize
    © 2025 IEEE. All rights reserved, including rights for text and data mining and training of artificial intelligence and similar technologies. Personal use is permitted, but republication/redistribution requires IEEE permission. See \href{https://www.ieee.org/publications/rights/index.html}{https://www.ieee.org/publications/rights/index.html} for more information.
  }%
}

\AddToShipoutPictureBG{%
  \AtPageUpperLeft{%
    \raisebox{-12pt}[0pt][0pt]{\makebox[\paperwidth]{\ieeetopheader}}
  }%
}

\AddToShipoutPictureBG{%
  \AtPageLowerLeft{%
    \raisebox{12pt}[0pt][0pt]{\makebox[\paperwidth]{\ieeebottomfooter}}
  }%
}

\begin{abstract}
Reducing a set of numbers to a single value is a fundamental operation in applications such as signal processing, data compression, scientific computing, and neural networks. Accumulation, which involves summing a dataset to obtain a single result, is crucial for these tasks. Due to hardware constraints, large vectors or matrices often cannot be fully stored in memory and must be read sequentially, one item per clock cycle. For high-speed inputs, such as rapidly arriving floating-point numbers, pipelined adders are necessary to maintain performance. However, pipelining introduces multiple intermediate sums and requires delays between back-to-back datasets unless their processing is overlapped. In this paper, we present JugglePAC, a novel accumulation circuit designed to address these challenges. JugglePAC operates quickly, is area-efficient, and features a fully pipelined design. It effectively manages back-to-back variable-length datasets while consistently producing results in the correct input order. Compared to the state-of-the-art, JugglePAC achieves higher throughput and reduces area complexity, offering significant improvements in performance and efficiency.

\end{abstract}


\vspace{-1.45em}
\section{Introduction}
\vspace{-0.2em}
In the realm of modern computing, the ability to efficiently reduce a set of numbers to a single value is fundamental to a wide variety of computational applications. This process, known as reduction, is crucial for tasks ranging from signal processing \cite{1ma, 2sarma}, data compression \cite{3jeong, 4rasheedi}, scientific computing \cite{5zhang, 6lee}, and neural networks \cite{7nahmias, 8wang}. 
Among reduction operations, accumulation, which sums a dataset into a single result, is one of the most fundamental. As data complexity and scale increase, high-performance accumulation methods that handle large datasets efficiently become essential.

Accumulation operations can be applied to both floating-point and integer data. For integer data, the use of a 3:2 compressor simplifies the process by reducing latency, making integer accumulation straightforward. However, floating-point accumulation presents more complex challenges, necessitating pipelined adders to manage high data input rates. Pipelining, while essential for maintaining throughput in high-speed applications, introduces design complexities such as managing multiple intermediate sums and potential delays with consecutive datasets.

\setlength{\tabcolsep}{4pt}

\setlength{\extrarowheight}{0.02cm}
\begin{table}[t]
    \centering
    \caption{Accumulation schedule for SimplePAC versus JugglePAC.}
    \vspace{-.75em}
    \scalebox{1}{
    \begin{tabular}{|c|c|c|c|c??c|c|c|c|} 
\hline

&\multicolumn{4}{|c??}{SimplePAC} & \multicolumn{4}{c|}{JugglePAC}  \\
\hline
\multirow{2}{*}{Cycle}&\multirow{2}{*}{Input} &  \multicolumn{3}{c??}{Adder} &  \multirow{2}{*}{Input} &\multicolumn{3}{c|}{Adder} \\ 

\cline{3-5}
\cline{7-9}
& & in1 & in2 & out & & in1 & in2 & out\\ 

\hline
\hline
0&$a_0$ & $a_0$ & 0& ~  &  $a_0$ && &~ 
\\ \hline
1&$a_1$ & $a_1$ & 0& ~  &  $a_1$ &$a_0$ & $a_1$ &~ 
\\ \hline
2&$a_2$ & $a_2$ & 0& ~  &  $a_2$ && &~ 
\\ \hline
3&$a_3$ & $a_3$& $a_0$& $a_0$  &  $a_3$ &$a_2$ & $a_3$ &
\\ \hline
4&$a_4$ & $a_4$& $a_1$& $a_1$  &  $a_4$ && &$a_{0,1}$
\\ \hline

5&$a_5$ & $a_5$& $a_2$& $a_2$  &  $a_5$ &$a_4$ & $a_5$ &
\\ \hline  
6&\multirow{5}{*}{Stall}& & & $a_{0,3}$ &  $b_{0}$
&$a_{0,1}$
& $a_{2,3}$
&$a_{2,3}$
\\ \cline{1-1} \cline{3-9} 
7& & $a_{0,3}$ & $a_{1,4}$ & $a_{1,4}$  &  $b_{1}$
&$b_{0}$
& $b_{1}$
&\\ \cline{1-1} \cline{3-9} 
8& & & & $a_{2,5}$ &  $b_{2}$&& &$a_{4,5}$\\ \cline{1-1} \cline{3-9}
9& & & &  &  $b_{3}$&$b_{2}$& $b_{3}$&$a_{0:3}$\\ \cline{1-1} \cline{3-9}
10& & $a_{2,5}$&                     $a_{0,1,3,4}$& $a_{0,1,3,4}$ &  $b_{4}$&$a_{4,5}$& $a_{0:3}$&$b_{0,1}$\\ \hline  
11& $b_{0}$&                  $b_{0}$&                     0&                      &  $b_{5}$&
$b_{4}$& $b_{5}$&\\ \hline
12& $b_{1}$& $b_{1}$ & 0&  &  $b_{6}$&$b_{0,1}$& $b_{2,3}$&$b_{2,3}$\\ \hline
13& $b_{2}$& $b_{2}$& 0 & $\mathbf{a_{0:5}}$ &  $b_{7}$&$b_{6}$& $b_{7}$ & $\mathbf{a_{0:5}}$\\ \hline
    \end{tabular}
    }
    \vspace{-2.1em}
    \label{table:schedule}
\end{table}

Table \ref{table:schedule} depicts an example of a floating-point accumulator using a single adder with a pipeline latency of 3 clock cycles, where new input values are fed in each cycle. Different datasets are represented by distinct letters, such as $a$ and $b$, with subscripts indicating the input indexes. For example, $a_0$ refers to the first element in dataset $a$, and $a_{0,1}$ denotes the sum $a_0 + a_1$.

In an accumulation circuit, the adder’s output feeds back to the corresponding new input. For SimplePAC, due to the adder’s pipeline latency of 3 clock cycles and the unavailability of immediate outputs, the first 3 inputs are added with zeros to initialize the accumulation. Once the outputs (subsums) are ready, they are summed with the next input, either from the fed input or previous subsums. In contrast, JugglePAC eliminates the need for zero-padding during the initial stage. Instead, when the first and second inputs are ready, they are immediately summed together. For example, in cycle 1, inputs $a_{0}$ and $a_{1}$ are summed without any zero-padding.

In SimplePAC, subsums are accumulated using a fixed reduction tree with P variables, where P is the adder's pipeline depth. However, this approach does not handle consecutive datasets properly, as results from datasets $a$ and $b$ may mix, leading to incorrect outcomes. While introducing stalls between datasets prevents this, JugglePAC improves upon this by alternating between datasets, ensuring the accumulation circuit stays fully pipelined without any stalls, as shown in Table \ref{table:schedule}.

This work presents JugglePAC, a novel fully pipelined accumulation circuit that optimizes both performance and area-timing efficiency using a single floating-point adder. 
The major contributions of this work are:

\begin{enumerate}
\item We propose JugglePAC, a fully pipelined accumulation circuit with a novel dynamic scheduling mechanism that efficiently handles high-speed, back-to-back variable-length datasets, overcoming control logic challenges and improving both performance and area efficiency.

\item Our design introduces innovative operation scheduling and dataset identification using labels, simplifying control logic, reducing area, and improving critical path performance in supported applications.

\item We implement JugglePAC across multiple target FPGAs, specifically the Xilinx XC2VP30 and XC5VLX50T. JugglePAC achieves the highest clock frequency for label widths 1, 2, and 3, reaching 208 MHz for the XC2VP30 and 334 MHz for the XC5VLX50T, outperforming existing solutions.

\item JugglePAC achieves competitive results in terms of area-timing efficiency (slices $\times$ $\mu$s) across label widths 1, 2, and 3, surpassing state-of-the-art solutions. As the label width increases, area-timing results increase. Even for label width 3 on the XC2VP30, where the result is 2,523, and label width 2 on the XC5VLX50T, with a result of 419, our design still outperforms other solutions.

\end{enumerate}

\vspace{-1.25em}
\section{Related Work}\label{Previous_work}
\vspace{-0.15em}
Recent work on floating-point accumulation circuits has aimed to optimize area, performance, and complexity. Early designs, such as Luo and Martonosi~\cite{luo2000accelerating}, used carry-save arithmetic and delayed adders but lacked full pipelining, causing performance stalls. Vangal~\cite{Vangal_2006} introduced pipelined multiply-accumulate units, though managing pipeline complexity remained challenging. He et al.~\cite{he2006group} improved accuracy using a group alignment algorithm but struggled with scalability for variable-length datasets.

Nagar and Bakos \cite{Nagar_2009} developed a double-precision accumulator with a coalescing reduction circuit, reducing complexity but limited by its FPGA-specific design. Zhou et al. \cite{zhuo2007high} introduced several designs, including the Fully Compacted Binary Tree (FCBT) and Dual Striped Adder (DSA), which managed multiple input sets but faced issues with buffer management and clock speed.

Huang and Andrews \cite{huang2012modular} proposed modular, fully pipelined architectures capable of handling arbitrary dataset sizes, though their designs faced challenges with underutilized pipelined adders and extensive buffering.

Recent designs \cite{sun2009floating, bachir2010performing, tai2009improved, tai2010multiple, tai2011accelerating} aimed at balancing area and timing performance. \cite{tai2011accelerating} notably improved the area-timing product but required multiple BRAMs, impacting area efficiency.

In contrast, JugglePAC offers a novel approach with a fully pipelined architecture that simplifies control logic and handles variable-length datasets efficiently. Its dynamic scheduling mechanism ensures high frequency and low area complexity, outperforming previous designs in scalability and adaptability across different FPGA architectures.

\vspace{-1.05em}
\section{JugglePAC} \label{Proposed_work}
JugglePAC is a novel floating-point accumulator optimized for performance and area. This section details its microarchitecture, inter-dataset behavior, and challenges like mixing variable-length datasets.

\subsection{JugglePAC Microarchitecture}

\begin{figure}[t]
\centering
\includegraphics[width=0.9\columnwidth]{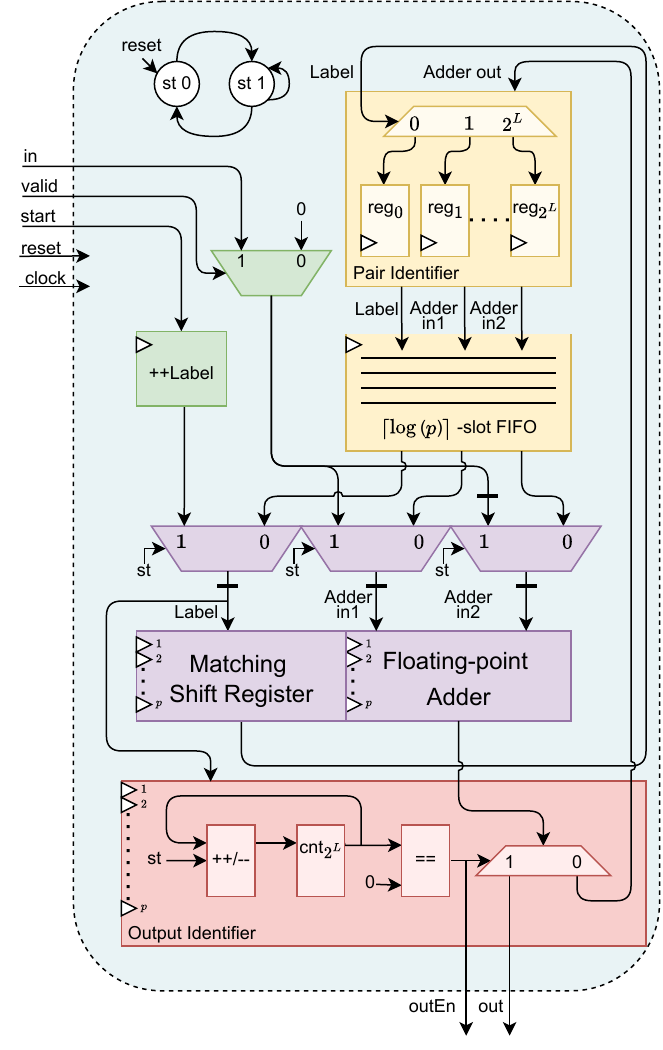}
\vspace{-1.35em}
\caption{JugglePAC's microarchitecture includes a Floating-Point Adder and key components such as a state machine for addition management, a Matching Shift Register for subsum labeling, and a Pair Identifier for scheduling. A FIFO buffer ensures smooth data flow, preventing congestion, while an Output Identifier tracks addition counts. This design enhances throughput while minimizing area complexity.}

\vspace{-0.65cm}
\label{fig:JugglePAC Architecture}
\end{figure}
The JugglePAC architecture is centered around a floating-point adder, which is critical for the accumulation process. It schedules serially arriving data to ensure efficient addition, operating at a throughput of 1 and performing additions every 2 cycles, resulting in 50\% utilization of the adder. This setup allows a single adder to keep up with the input rate when scheduling is managed effectively. JugglePAC employs a state machine with two states: in the state 1, inputs are directly added, while in state 0, available subsums are accumulated, as illustrated in Fig. \ref{fig:JugglePAC Architecture}. The state machine alternates between these states each cycle, except for datasets with an odd length, where it remains in state 1 for an additional cycle. 

To enhance performance in high-throughput applications, JugglePAC processes back-to-back inputs, preventing data pile-ups. This requires handling subsums from multiple datasets simultaneously. Each subsum is assigned a unique incrementing label (green in Fig. \ref{fig:JugglePAC Architecture}) to differentiate datasets.   
A shift register, with latency matching the adder (purple in Fig. \ref{fig:JugglePAC Architecture}), maintains this labeling. The Matching Shift Register ensures subsums are correctly paired within the same dataset while distinguishing those from different datasets. To enhance throughput, additional pipeline stages are introduced before the adder inputs. The registers corresponding to these pipeline stages are depicted by short thick lines in Fig. \ref{fig:JugglePAC Architecture}, placed along the connections between different blocks. Depending on the desired application, intentional gaps or delays may occur within or between datasets, during which the valid input is set to 0. In these cases, the green multiplexer in Fig. \ref{fig:JugglePAC Architecture} forces the input to 0 during accumulation, ensuring the output remains unaffected.

Scheduling the additions of subsums requires efficient control logic. The Pair Identifier (PI) block, shown in Fig. \ref{fig:JugglePAC Architecture} and highlighted in yellow, manages this process. It stores incoming subsums in registers, identifies addition pairs, and schedules the operations accordingly. The number of registers depends on the label size (L). To handle periods when the adder is unavailable, JugglePAC uses a FIFO buffer to store ready-to-add subsum pairs along with their labels. This approach allows efficient juggling between datasets while minimizing area requirements. 
After the last input is received, the circuit holds at most $p$ subsums in the pipeline, which require $\log(p)$ cycles to reduce. Thus, the next accumulation needs to buffer at most log(p) values. Consequently, the FIFO has a depth of $\lceil \log(p) \rceil$, where $p$ is the adder's latency. The FIFO's design, which utilizes registers, ensures efficient data management.

The state machine in JugglePAC controls the accumulation process with low area and timing complexity. It alternates between adding serial input data and processing FIFO data, managed by the purple-colored multiplexers. Table~\ref{table:schedule} shows a sample schedule, where accumulation of dataset $b_{0:N}$ begins before the completion of $a_{0:5}$. The simple state machine and pair identification logic help JugglePAC achieve low area and critical path delays, outperforming state-of-the-art designs

The simplest way to identify when a dataset’s final output is obtained is by counting inputs and additions. For $N$ values, $N-1$ additions are required; however, $N$ is not fixed. An up-down counter handles this by incrementing for each input and decrementing for each addition. JugglePAC optimizes this by tracking the number of additions performed from each state. The output identifier is shown in Fig. \ref{fig:JugglePAC Architecture}, color-coded in red. The counter increments for each addition from state 1 and decrements for each addition from state 0, skipping the first addition of state 1 to reset after each operation.

For the first $p+3+(1-(p \bmod 2))$ cycles (where $p$ is the adder’s latency), only state 1 performs additions. The term $p + 3$ accounts for the cycles required to generate two subsums, while the additional $1 - (p \bmod 2)$ reflects the scheduling delay before state 0 can perform its first addition. After this period, state 1 continues to perform additions more frequently than state 0 until a steady state is reached (when both states perform additions at the same rate). The counter reaches a maximum of $p + 2$, representing the buffer size before state 0 schedules its next addition (i.e., before the FIFO). The additional 2 cycles in this value account for the buffers in the Pair Identifier block and a pipeline register between the FIFO and the adder.

\vspace{-1.3em}
\subsection{Inter-Dataset Behavior}
\vspace{-0em}
The JugglePAC architecture allows for parallel processing of datasets, with the label size determining the maximum number of datasets that can be handled simultaneously. When the number of datasets exceeds $2^L$, different datasets may be assigned the same label, causing mixing and incorrect results. However, if the dataset lengths are long enough, this issue is avoided. To ensure this, JugglePAC introduces a minimum dataset length constraint, which depends on the label size.

To calculate the minimum dataset length, we consider the maximum overlap between consecutive datasets, which defines this lower bound. Starting from the simple case where the label size is 1, only two datasets can overlap. Each dataset can have at most $p$ subsums in the system after its last input is received, where $p$ is the adder's latency, resulting in $\lceil\log p\rceil$ levels in the binary reduction tree. The output identification logic excludes the initial addition from state 1, which effectively increases the scheduling latency by an additional $p$ cycles. Additionally, JugglePAC requires datasets to be larger than 4 to be recognized due to the output identification logic, which is accounted for as a constant offset and a minimum. As the label size increases, more overlap is possible. Using the base case, the minimum dataset length is calculated using \eqref{eq:dataset_ceil}.

\vspace{-0.5em}
\begin{equation}
\max( \left\lceil \frac{ (1 + \lceil\log p\rceil) p + 4}{2^L - 1} \right\rceil, 4)
\label{eq:dataset_ceil}
\end{equation}

JugglePAC’s dynamic accumulation approach leads to latency variations based on the current and previous dataset lengths. For label widths less than 3, latency stays consistent, with variations under 5 cycles, maintaining input order. For larger widths, short datasets can cause results to appear out of order. This can be corrected using control logic to reorder outputs based on the label or by outputting the label itself. Our analysis shows that setting the minimum dataset length to 19 consistently preserves input order across all test cases. This design effectively balances high frequency and low area complexity, outperforming current state-of-the-art solutions.

\vspace{-1em}
\section{Implementation Results} \label{Results}
\vspace{-.1em}
We evaluated the JugglePAC floating-point accumulation circuit against prior designs, focusing on area, latency, throughput, and frequency. Table~\ref{tab:table2} summarizes these metrics, with dashes (--) indicating unavailable or inapplicable values. Notably, the label width parameter is unique to JugglePAC; since the minimum dataset length depends on it, this metric is not relevant to previous works.

\setlength{\tabcolsep}{4pt}

\begin{table*}[t]
    \centering
    \caption{Comparison with previously proposed accumulation circuits. }
    \vspace{-1em}
    \begin{tabular}{|c |c c c c c c c c c c c| c|}
 \hline 
   & & & & & & & & & & & & \\ [-0.3cm]
 \multirow{2}{*}{Design} &  Label &Min. Dataset & \multirow{2}{*}{Adders} & \multirow{2}{*}{Slices}& \multirow{2}{*}{Registers}& \multirow{2}{*}{LUTs} & \multirow{2}{*}{BRAMs} & Frequency  & \multicolumn{2}{c}{Total Latency}  & \multirow{2}{*}{Slices$\times \mu$s} & \multirow{2}{*}{FPGA}\\ [0.4ex] 
  
  &   Width&Length& & & & & & (MHz)& cycles & $\mu$s & &\\

 \hline\hline
  MFPA \cite{huang2012modular} &  - &- & 4 & 4,991 & -& -& 2  & 207 & 198 & 0.957 & 4,776 &\multirow{11}{*}{XC2VP30} \\ 
 \cline{1-12}  
 
 AeMFPA \cite{huang2012modular} &  - &- & 2 & 3,130 &- &- & 14 & 204 & 198 & 0.970& 3,036 &\\ 
 \cline{1-12}
 
 $Ae^{2}$MFPA \cite{huang2012modular} &  - &- & 2 & 3,737 & -&- & 2 & 144 & 198 & 1.370& 5,120 &\\ 
 \cline{1-12}
 
  FAAC \cite{sun2009floating} &  - &- & 3 & 6,252 &- &- & 0 & 199 & 176 & 1.086 & 6,790 &\\ 
 \cline{1-12}
 
 FCBT \cite{zhuo2007high} &  - &- & 2 & 2,859 & -&- & 10 & 170 & $\le$ 475 & $\le$ 2.794 & 7,988 &\\ 
 \cline{1-12}
 
 DSA \cite{zhuo2007high} &  - &- & 2 & 2,215 & - &- & 3 & 142 & 232 & 1.634 & 3,619 &\\ 
 \cline{1-12}
 
 SSA \cite{zhuo2007high} &  - &- & 1 & 1,804 & - & - & 6 & 165 & $\le$ 520 & $\le$ 3.152 & 5,686 &\\ 
 \cline{1-12}
 
 DB \cite{tai2011accelerating}  &  - &- & 1 & 1,749 &- &- & 6 & 188 & $\le$ 199 & $\le$ 1.058 & 1850 &\\  \cline{1-12}

\textbf{JugglePAC}  &  1&74 & 1 & 1,439 & 1763 & 2010 & 0 & 208 & $\le$ 220& $\le$ 1.058& 1,522& \\ \cline{1-12} 

 \textbf{JugglePAC} &  2&25 & 1 & 1,796 &1910 &2592  & 0 & 208 & $\le$ 224& $\le$ 1.077& 1,934 & \\  \cline{1-12} 

  \textbf{JugglePAC}  &  3&11 & 1 & 2,343 & 2199 & 3418  & 0 & 208 & $\le$ 224& $\le$ 1.077& 2,523  & \\  \cline{1-12} 
  \textbf{JugglePAC}  &  4&  5 & 1 & 3,339 & 2800 & 5086 & 0 & 159 & $\le$ 224& $\le$ 1.409 & 4,704  & \\  \hline 
  \hline
  FPACC \cite{bachir2010performing}  & - & - & -  & 683 & 1744 &2234  &-& 247 & - & - & - & VC5VSX50T\\ 
  \cline{1-13}
  \textbf{JugglePAC} &2&25&1& 625 & 1702 & 1607 & 0& 334 & $\le$224 & $\le$ 0.671& 419 & XC5VLX50T\\ \hline
  BTTP \cite{tang2021novel} & - & - & 1 &  648& 1554 & 1495 & 9.5 & 305 & - & - & - & XC5VLX110T\\ 
  \hline
\end{tabular}
    \label{tab:table2}
    \vspace{-1.8em}
\end{table*}
\vspace{-0.25em}

JugglePAC demonstrates a significant reduction in area complexity, using fewer slices than designs like MFPA, AeMFPA, and FAAC \cite{huang2012modular}, with up to 71\% less slice usage. Additionally, JugglePAC operates without BRAMs, unlike FCBT and DSA \cite{zhuo2007high}, contributing to its lower area complexity. 

In terms of latency, JugglePAC performs competitively. For a label width of 2 and minimum dataset length of 25, JugglePAC achieves a latency of approximately 1.077 $\mu$s, comparable to or better than most previous designs such as DSA and SSA \cite{zhuo2007high}. Its throughput is also high, outperforming designs like DB \cite{tai2011accelerating} and BTTP \cite{tang2021novel}, especially with larger datasets. The reported latencies represent the worst-case execution time as the latency is dependent on the lengths of consecutive datasets as well.

JugglePAC operates at a frequency of 208 MHz when the label width is 1, 2, or 3, surpassing many previous designs like FPACC \cite{bachir2010performing} and FCBT \cite{zhuo2007high}. It also achieves the lowest "slices $\times$ $\mu$s" score for these label widths, reflecting its superior efficiency in balancing area and performance. However, as the label width increases, the area-timing (slices $\times$ $\mu$s) rises. Thus, our design is best suited for applications that require a small label width.

JugglePAC's full pipelining improves performance and resource utilization, excelling in high-speed operations. However, using a single floating-point adder may limit performance in multi-adder scenarios. Its minimum dataset length varies with label width, ensuring accurate accumulation but potentially limiting flexibility. Future work could explore using multiple adders to overcome these limitations and boost performance.

\vspace{-1em}
\section{Conclusion}\label{Conclusion}
\vspace{-.1em}
Accumulation is a fundamental operation found in many computational workloads. High-throughput floating-point accumulation presents challenges due to pipelining, especially when handling consecutive datasets of varying lengths and preserving input order. Existing approaches often incur significant overhead, leading to reduced clock frequency or increased area complexity. This work introduced JugglePAC, a novel, fully pipelined reduction circuit that uses a single floating-point adder to manage accumulation efficiently. Evaluated on multiple FPGAs, JugglePAC consistently achieves both low area and high timing performance, while prior designs typically improve one but not both at the same time. 

\vspace{-1.25em}

\vspace{-0.15em}
\bibliographystyle{IEEEtran}

\bibliography{output.bib}

\end{document}